\documentclass[superscriptaddress,secnumarabic,
amssymb,amsmath,nobibnotes,aps,prd,showkeys,nofootinbib]{revtex4}

\usepackage{amsmath}
\usepackage{amsfonts}
\usepackage{amsthm}
\usepackage{amssymb}
\usepackage{amscd}
\usepackage[british]{babel}
\usepackage{graphicx}
\usepackage{psfrag}
\usepackage{epsfig}
\usepackage{rotating}
\usepackage{natbib}%
\usepackage{times}

\usepackage{ulem}

\usepackage{color}
\usepackage{float}
\usepackage{url}
\usepackage[pdftex, colorlinks=true, citecolor=red, linkcolor=blue, urlcolor=black]{hyperref}
\hypersetup{linktocpage}

\begin{document}

\title{Qualitative study in Loop Quantum Cosmology}

\author{Llibert Arest\'e Sal\'o}
\email{llibert.areste@estudiant.upc.edu}

\author{Jaume Amor\'os}
\email{jaume.amoros@upc.edu}
\author{Jaume de Haro}
\email{jaime.haro@upc.edu}
\affiliation{Departament de Matem\`atica Aplicada, Universitat Polit\`ecnica de Catalunya, Diagonal 647, 08028 Barcelona, Spain}

\begin{abstract}

This work contains a detailed qualitative analysis, in General Relativity and in Loop Quantum Cosmology, of the dynamics in the associated phase space of a scalar field minimally coupled with gravity, whose potential
mimics the dynamics of a perfect fluid with a linear Equation of State (EoS). 
Dealing with the orbits (solutions) of the system, we will see that
there are  analytic ones, which lead to the same dynamics as the perfect fluid, and our goal is to 
check their stability, depending on the value of the EoS parameter, i.e., to show whether the other orbits converge or diverge to these analytic solutions at early and late times.

\end{abstract}

\maketitle




\section{Introduction}

Soon after Einstein published the field equations of General Relativity  (GR), it was realized that these equations contained singularities. In particular, in a cosmological context, it was noticed that for the Friedmann-Lema{\^\i}tre-Robertson-Walker geometry, when the Equation of State (EoS) modeling the matter content was a linear equation with an EoS parameter greater than $-1$, a singularity named Big Bang appeared at early times, where the energy density of the universe diverges. Moreover, dealing with nonlinear EoS
(see for instance, \cite{leonardo1}, where the analysis is done by expressing the EoS in the form of a power expansion), one can see that other kind of singularities such as Sudden singularity \cite{barrow1,nojiri1,barrow2} or Big Freeze \cite{nojiri2,bouhmadi,nojiri3, nojiri5, nojiri6} appear.

\

Several attempts to remove these kinds of singularities were studied for years. We can mention, for instance, the assumption of a non-linear EoS \cite{telleparallel} with two fixed points where the pressure plus energy density vanishes, or the modification of the Friedmann equation assuming that the square of the Hubble parameter is equal to a power series of the energy density \cite{leonardo2}.
Another option is the introduction of higher order terms in the equations of GR via the quantum effects of a scalar field conformally coupled with gravity \cite{starobinsky,fischetti,elizalde1,elizalde2,elizalde3} or directly assuming a Lagrangian containing non-linear terms in the scalar curvature \cite{amendola,nojiri4,bamba}. However, the simplest way to remove the Big Bang singularity is to adopt the viewpoint of Loop Quantum Cosmology (LQC), where the discrete nature of spacetime is assumed and Friedmann equations are modified  (see for review \cite{singh1,singh2}). Hence, all strong curvature singularities are resolved in LQC, both in isotropic \cite{singh3,singh4} and anisotropic models \cite{singh5,singh6} (see \cite{singh7} for a complete review). Thus, several types of singularities are avoided (for e.g. Big Rip \cite{singh8} and Big Crunch \cite{singh9}) and the Big Bang is replaced by a non-singular bounce (see for instance \cite{singh10,singh11}).

\

However, a qualitative analysis of the dynamics, both in GR and LQC, is only done in few works  \cite{heard, sami, naskar, samart}. For this reason, the aim of this paper is to perform a qualitative analysis in the spatially flat FLRW, for both GR and  Loop Quantum Cosmology (LQC), when the universe is filled 
with a scalar field whose potential leads to a solution conducing to a dynamics which mimics the well-known perfect fluid with EoS $P=\omega\rho$. More precisely, 
we will qualitatively study when we have a scalar field $\varphi$ which mimics this fluid by solving the conservation equation with the effective potential to which this scalar field is submitted. Thus, since when one deals with an scalar field the conservation equation is a second order differential equation,  we obtain an infinite set of orbits which can be plotted in the phase space $(\varphi,\dot{\varphi})$ and some of which may not satisfy the equation of state, i.e., some of them do not depict a universe with a constant EoS parameter, meaning that they do not mimic the same background as the perfect fluid. Then, the question is whether these orbits depicts at early or late time the same background as the perfect fluid. With our analysis we can infer whether the orbits with the same background as the fluid are attractors or repellers of the dynamical system, in the sense that the other orbits starting in their neighbourhood asymptotically approach them or not. This study is very important in the so-called matter bounce and matter-ekpyrotic bounce scenario 
\cite{WilsonEwing:2012pu,Wilson-Ewing:2013bla,Cai:2014zga,Haro:2013bea} (recall that singularities  in  ekpyrotic LQC was studied for the first time in  \cite{singh9, singh12, singh13}) because all the calculations -the power spectrum of perturbations, the spectral index, its running \cite{Lehners:2015mra,WilsonEwing:2012pu,Elizalde:2014uba} and the reheating temperature \cite{Haro:2015zda,Quintin:2014oea}-
are done with the analytical solution, but of course these calculations are only viable if this solution is stable, in the sense that all the other solutions depict asymptotically a universe with the same properties.

\

The work is organized as follows: 

In section \ref{einstein}, we introduce the Friedmann equations for a flat homogeneous and isotropic space-time and we reconstruct the scalar field and the corresponding potential which leads to the same background as the fluid with EoS $P=w\rho$. Using this potential, we perform a qualitative study of the dynamical equations, obtaining the phase portraits for the different values of the effective EoS parameter, and concluding that when the effective EoS parameter belongs to the interval $(-1,1)$ the analytical orbit is a repeller in the contracting phase and an attractor in the expanding one. On the contrary, when the effective EoS parameter is greater than $1$, i.e. in the ekpyrotic case, the analytical orbit is an attractor in the contracting phase and a repeller in the expanding one. 

In section \ref{lqc}, we  proceed analogously as with GR, but now including in the Friedmann equations holonomic corrections that come from LQC \cite{bambaharo}. 

Finally, we do as well a qualitative study for the dynamical equations of the scalar field, obtaining similar phase portraits as the one shown in \cite{faseharo} for a particular case of the ones we are going to study in the present work.

\

We will use natural units ($\hbar=8\pi G=c=1$).

\

\

\section{General Relativity} \label{einstein}

\subsection{Friedmann Equations}

Assuming homogeneity and isotropy of the universe, which is valid at sufficiently large scales, it can be shown that a suitable change of coordinates leads to the so-called Friedmann-Lema\^itre-Robertson-Walker (FLRW) metric
\begin{equation}
ds^2=dt^2-a^2(t)\left(\frac{dr^2}{1-\kappa r^2}+r^2d\theta^2+r^2\sin^2\theta d\varphi^2\right),
\end{equation}
where $a(t)$ is a scale factor that parametrizes the relative expansion of the universe and the curvature $\kappa$ is -1,0 or 1 if we are dealing respectively with an open, flat or closed universe.

\

To simplify, 
we will perform all our calculations in the flat FLRW space-time. Hence, with the EoS $P=w\rho$, we obtain the Friedmann equations
\begin{equation}
        H^2=\frac{\rho}{3}, \ \ \ \ \ \ \ \ \ \ \ \ \  \dot{H}=-\frac{3(1+w)}{2}H^2, \ \ \ \ \ \ \ \ \ \ \ \ \
       \dot{\rho}=-3H(1+w)\rho, \label{friedmann}
\end{equation}
where $H=\frac{\dot a}{a}$ is the Hubble parameter.

\

The dynamics provided by (\ref{friedmann}) could be mimicked by a scalar field minimally coupled with gravity. Its potential,
which is exponential \cite{heard}, is easily obtained using the reconstruction method which goes as follows. For a minimally coupled with gravity scalar field, one has $\rho=\frac{\dot\varphi^2}{2}+V(\varphi)$ and $P=\frac{\dot\varphi^2}{2}-V(\varphi)$. Therefore, the linear EoS leads to
\begin{eqnarray}\label{A}
V(\varphi)=\frac{1-w}{1+w}\frac{\dot\varphi^2}{2}.\end{eqnarray}

On the other hand, solving the Friedmann equations one can show that for the linear EoS the energy density evolves as $\rho(t)=\frac{4}{3(1+w)^2t^2}$, which implies, using
 $\dot{\varphi}^2=P+\rho=(1+w)\rho$, that
 \begin{eqnarray}\label{B}
 \dot{\varphi}=\frac{2}{\sqrt{3(1+w)}}\frac{1}{|t|},
 \end{eqnarray}
where we have chosen, for simplicity, the positive sign of the square root. Integrating this equation, one obtains the following two solutions
\begin{eqnarray}
\varphi=\pm\frac{1}{\sqrt{3(1+w}}\ln \left( \frac{t}{t_0} \right)^2,
\end{eqnarray}
where the sign $+$ (resp. $-$) refers to a solution depicting a universe in the expanding (resp. contracting) phase. Finally, from this equation one deduces that 
$\left( \frac{t}{t_0} \right)^2=e^{\pm \sqrt{3(1+w)}\varphi}$. Then, using this expression and Equations \eqref{A} and \eqref{B}, one concludes that
\begin{eqnarray}
V_{\pm}(\varphi)=V_0e^{\mp\sqrt{3(1+\omega)}\varphi}, \label{potrg}
\end{eqnarray}
where $V_0=\frac{2(1-w)}{3(1+w)^2t_0^2}$ and  $V_+$ (resp. $V_-$) denotes  the potential for the expanding (resp. contracting) phase.


\subsection{Qualitative analysis in GR}

In the case of the linear EoS $P=\omega \rho$ with $\omega>-1$, we want to study the behaviour of the analytical solution in a similar way as done in \cite{heard}, finding out whether it is either an attractor or a repeller and comparing its behaviour with that of other solutions of the system. Using the potential found in \eqref{potrg}, we will analyse the dynamics in the contracting phase from this dynamical system
\begin{equation}
\ddot{\varphi}+3H_-(\varphi,\dot{\varphi})+V_{\varphi}=0,
\end{equation}
where $H_-(\varphi,\dot{\varphi})=-\sqrt{\frac{\dot{\varphi}^2}{2}+V(\varphi)}$.

\

With the change of variable $\varphi=\frac{-2}{\sqrt{3(1+\omega)}}\ln\psi$,
the dynamical system becomes
\begin{equation}
    \frac{d\dot{\psi}}{d\varphi}=F_{-}(\dot{\psi}):=-\frac{3}{2}\sqrt{1+\omega}\left(\sqrt{\frac{2\dot{\psi}^2}{3(1+\omega)}+V_0}+\frac{\sqrt{3}(1+\omega)}{2\dot{\psi}}\left(\frac{2\dot{\psi}^2}{3(1+\omega)}+ V_0 \right) \right). \label{dyneinstein}
    \end{equation}
    
The different cases to distinguish are the following ones:

\begin{itemize}
\item $\omega=1$: This case is known as a kination (or deflationary) phase \cite{joyce,spokoiny}. Equation \eqref{dyneinstein} becomes $\frac{d\dot{\psi}}{d\varphi}=-\sqrt{\frac{3}{2}}(|\dot{\psi}|+\dot{\psi})$ and 
in the semi-plane $\dot{\psi}\leq 0$ ($\dot{\varphi}\geq 0$) the solution is given by
\begin{equation}
(\varphi(t),\dot{\varphi}(t))=\left(-\sqrt{\frac{2}{3}}\ln(-|C|(t-t_s)),-\sqrt{\frac{2}{3}}\frac{1}{t-t_s}\right) \ \quad \mbox{with} \quad  t<t_s,
\end{equation}
which is an orbit that trivially fulfills the Equation of State because of the potential being null, with the sign corresponding to the contracting phase. Despite not being stable because of the singularity at $t=t_s$, all possible values of $C$ lead to solutions corresponding all the time to a universe with EoS $P=\omega \rho$ in the contracting phase, with $H(t)=\frac{1}{3(t-t_s)}$.

Regarding the semi-plane $\dot{\psi}>0$ ($\dot{\varphi}<0$), the solution becomes
\begin{equation}
(\varphi(t),\dot{\varphi}(t))=\left(\sqrt{\frac{2}{3}}\ln(-|C|(t-t_s)),\sqrt{\frac{2}{3}}\frac{1}{t-t_s}\right) \ \quad \mbox{with} \quad  t<t_s,
\end{equation}
which is analogous to the former case.

Finally, the case $\dot{\psi}=0$, i.e. $(\varphi(t),\dot{\varphi}(t))=(C,0)$, corresponds to $H=0$.

\

\item $-1<\omega <1$: Given that $F_-(\dot{\psi})$ can only vanish for $\dot{\psi}<0$, we have a single critical point for $\dot{\psi}$
\begin{equation}
\dot{\psi}_-=-(1+\omega)\sqrt{\frac{3V_0}{2(1-\omega)}},
\end{equation}
which is a global repeller, given that $F_-(\dot{\psi})>0 \ \ \forall \dot{\psi}_-<\dot{\psi}<0$ and $F_-(\dot{\psi})<0 \ \ \forall \dot{\psi}<\dot{\psi}_-$.

\begin{figure}[H]
\begin{center}
\includegraphics[height=35mm]{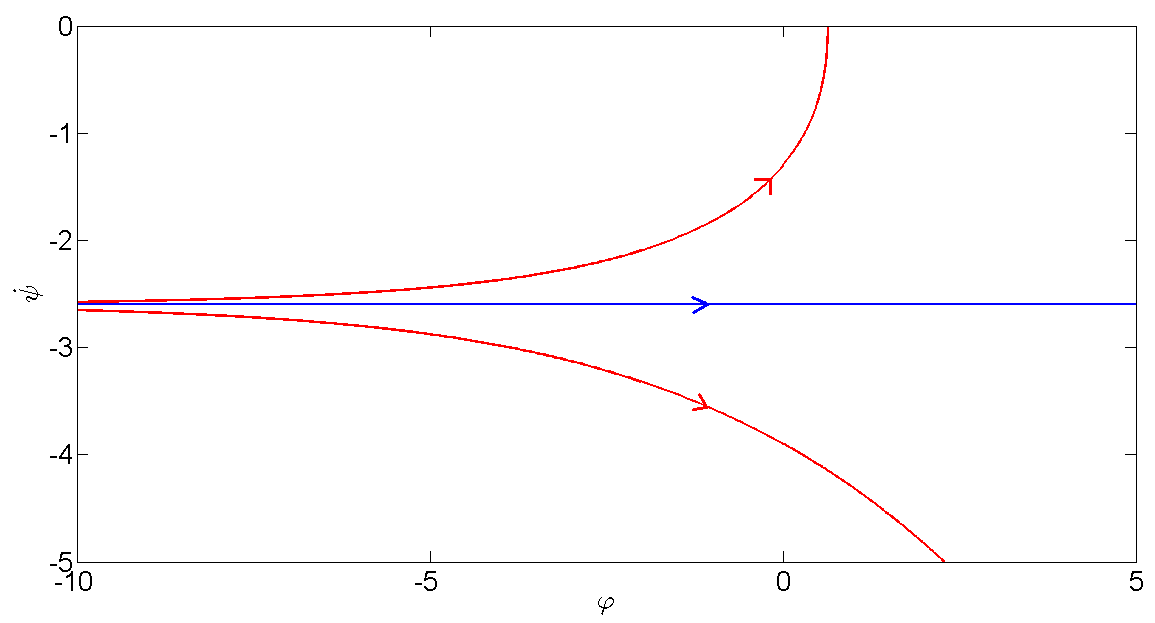}
\end{center}
\caption{Phase portrait in the phase space $(\varphi,\dot\psi)$ for $\omega=1/2$ and $V_0=1$. The blue orbit is the analytic one which mimics the orbit given by a perfect fluid with
EoS parameter $\omega=1/2$. The red orbits are obtained numerically to show that, in the contracting phase, the analytic orbit is a global repeller.}
\label{figgrwl1}
\end{figure}

We point out that it can be verified that the blue horizontal line in Figure \ref{figgrwl1} verifies the Equation of State, depicting a universe with EoS $P=\omega \rho$.

\

\item $\omega>1$: This case is known as an ekpyrotic phase or regime \cite{khoury} and the system is only defined for $|\dot{\psi}|\geq \sqrt{\frac{3(1+\omega)|V_0|}{2}}$. We have three critical points
\begin{equation}
\dot{\psi}_-=-(1+\omega)\sqrt{\frac{3|V_0|}{2|1-\omega|}} ,\ \ \ \ \ \ \ \dot{\psi}_0^{\pm}=\pm \sqrt{\frac{3(1+\omega)|V_0|}{2}},
\end{equation}
where
$\dot{\psi}_0^{\pm}$ are repellers corresponding to $H=0$. On the other hand, $\dot{\psi}_-$ is an attractor for $\dot{\psi}<\dot{\psi}_0^-$, solution that leads to a universe that all the time behaves as $P=\omega\rho$ in the contracting phase.

\begin{figure}[H]
\begin{center}
\includegraphics[height=40mm]{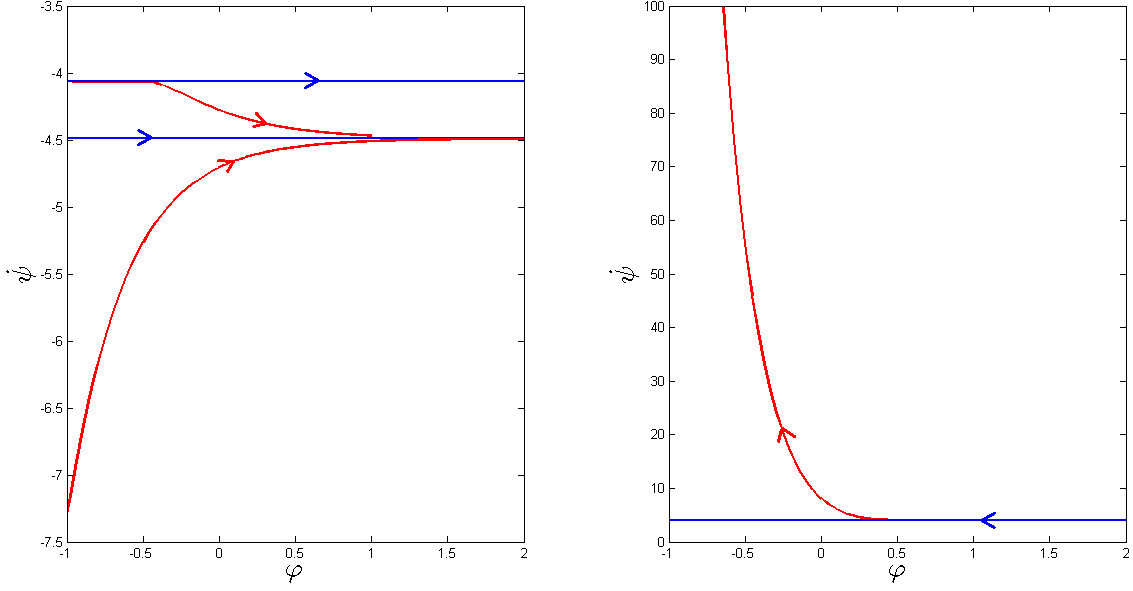}
\end{center}
\caption{Phase portrait in the phase space $(\varphi,\dot\psi)$ for $\omega=10$ and $V_0=-1$. In the left-hand side picture we have plotted in blue, in the semi-plane $\dot\psi<0$,  the orbit $\dot\psi^-$ which mimics the background given by a perfect fluid with EoS parameter $w=10$ , and one corresponding to $H=0$, i.e. $\dot{\psi}_0^-$. The red orbits obtained numerically show that 
$\dot\psi^-$ is an attractor and $\dot{\psi}_0^-$ is a repeller. On the right-hand side, in the semi-plane $\dot\psi>0$, we have plotted in blue the other orbit corresponding to $H=0$, i.e. $\dot{\psi}_0^+$, and in red an
orbit obtained numerically  which shows that  $\dot{\psi}_0^+$ is a repeller.}
\end{figure}

We note that orbits approaching either of the fixed points $\dot{\psi}_0^{\pm}$ reach such value in a finite cosmic time, which can be justified in the following way. Since near $\dot{\psi}_0^{\pm}$ the dynamical system behaves as $\frac{d\dot{\psi}}{d\varphi}=-\frac{3}{2}\sqrt{1+\omega}\sqrt{\frac{2\dot{\psi}^2}{3(1+\omega)}+V_0}$, it is reached in a finite $\varphi$ time. Hence, $\dot{\varphi}\left(=\sqrt{-2V(\varphi)}\right)$ is finite as well and, thus, in a finite cosmic backwards time the contracting phase starts after a bounce at $\rho=0$ enabling the system to leave the expanding phase with no singularity.

\

An analogous analysis for the expanding phase would show that for $\omega = 1$ there are as well solutions with $H=0$ and other non-trivial ones, all satisfying the corresponding Equation of State. The solution that depicts a fluid with EoS $P = \omega\rho$ is an attractor for $|\omega| < 1$ and a repeller for $\omega>1$, also verifying the Equation of State. And for $\omega>1$ we would also find critical points corresponding to $H=0$, as the ones found in the contracting phase, that are in this case attractors.

\end{itemize}

\

\section{Loop Quantum Cosmology} \label{lqc}

\subsection{Modified Friedmann equations}

The general formula of loop quantum gravity, which takes into account the discrete nature of space-time, expresses the Hamiltonian in terms of the holonomies $h_j(\lambda)\equiv e^{-i\frac{\lambda\beta}{2}\sigma_j}$, where $\sigma_j$ are the Pauli matrices \cite{bambaharo,singh}
\begin{equation}
\mathcal{H}_{\text{LQC}}=-\frac{2V}{\gamma^3\lambda^3}\sum\limits_{i,j,k}\epsilon^{ijk}\text{Tr}[h_i(\lambda)h_j(\lambda)h_i^{-1}(\lambda)h_j^{-1}(\lambda)\{h_k^{-1}(\lambda),V\}+\rho V,\label{ham}
\end{equation}
where $\gamma\approx 0.2375$ is the Barbero-Immirzi parameter \cite{meissner}, $\lambda=\sqrt{\frac{\sqrt{3}}{4}\gamma}$ is a parameter with the dimension of length, which is determined by invoking the quantum nature of the geometry (see for instance \cite{PSingh}), and $\beta$ is the canonically conjugate variable of the volume $V=a^3$ and whose Poisson bracket satisfies
$\{\beta, V\}=\frac{\gamma}{2}$, which means that at classical level, i.e. when $H$ is small and holonomy corrections can be disregarded, $\beta=\gamma H$.

\

The Hamiltonian expression in \eqref{ham} leads to \cite{piotr,elizalde4}
\begin{equation}
\mathcal{H}_{\text{LQC}}=-3V\frac{\sin^2(\lambda\beta)}{\gamma^2\lambda^2}+\rho V,
\end{equation}
and, by imposing the Hamiltonian constraint $\mathcal{H}_{\text{LQC}}=0$,  one obtains the following holonomy corrected 
Friedmann equation
\begin{eqnarray}\label{constraint}
\frac{\sin^2(\lambda\beta)}{\gamma^2\lambda^2}=\frac{\rho}{3}. \end{eqnarray}

On the other hand, 
the Hamiltonian equation 
$\dot{V}=\{V,\mathcal{H}_{\text{LQC}}\}=-\frac{\gamma}{2}\frac{\partial\mathcal{H}_{\text{LQC}}}{\partial\beta}$
leads to 
\begin{eqnarray}
H=\frac{\sin 2\lambda \beta }{2 \lambda\gamma} \Longleftrightarrow \beta=\frac{1}{2\lambda}\arcsin (2\lambda\gamma H).
\end{eqnarray}

Inserting the value of $\beta$ in the Hamiltonian constraint \eqref{constraint}, one obtains
\begin{eqnarray}
\frac{\sin^2\left(\frac{1}{2}\arcsin (2\lambda\gamma H)  \right)}{\gamma^2\lambda^2}=\frac{\rho}{3}. \end{eqnarray}

Note that the geometric corrections modify the geometric sector of the Friedmann equation and for small values of the Hubble parameter one recovers its usual form in General Relativity.

\

However, using the relation $\sin^2(\frac{x}{2})=\frac{1-\cos x}{2}$ and after some algebra, this equation becomes
\begin{eqnarray}
\frac{1-\sqrt{1-4\lambda^2\gamma^2 H^2}}{2\lambda^2\gamma^2}=\frac{\rho}{3},
\end{eqnarray}
and finally isolating $H^2$ one obtains the more usual form
\begin{equation}
H^2=\frac{\rho}{3}\left(1-\frac{\rho}{\rho_c}\right), \label{ellipse}
\end{equation}
where $\rho_c=\frac{3}{\gamma^2\lambda^2}$ is the so-called critical energy density (the maximum value that reaches the energy density).

\

{\bf Remark:} It is important to note that the effective Friedmann equation (\ref{ellipse}) of LQC has been  fully trusted only for some values of the parameter $\omega$. In particular, in \cite{taveras} it was verified for $\omega=1$.


\

Equation \eqref{ellipse} corresponds to an ellipse in the plane $(\rho, H)$, that we can parametrize in the following form
\begin{equation}
\left\{ \begin{array}{ll} H=\sqrt{\frac{\rho_c}{12}} \sin\eta  \\   \rho=\rho_c \cos^2\frac{\eta}{2}. \end{array} \right. \label{param}
\end{equation}

The conservation equation does not differ from standard GR, i.e.,   the fluid fulfills the relation $d(\rho V)=-PdV$, where $V=a^3$, and  again with the EoS $P=-\rho-f(\rho)$ one gets $\dot{\rho}=3Hf(\rho)$. Hence, we easily obtain the Raychaudhuri equation
$$\dot{H}=\frac{f(\rho)}{2}\left(1-\frac{2\rho}{\rho_c}\right).$$

\

Analogously as in the case of General Relativity, one can mimic the perfect fluid that recovers the space-time with a scalar field and one can verify that the potential to which this scalar field is submitted is

\begin{equation}
V(\varphi)=V_0\frac{e^{\sqrt{3(1+\omega)}\varphi}}{\left(1+\frac{V_0}{2\rho_c(1-\omega)}e^{\sqrt{3(1+\omega)}\varphi} \right)^2}. \label{potalfa1}
\end{equation}

\

{\bf Remark:} This potential could be obtained using the reconstruction method as we have done in the case of GR or, as in \cite{mielczarek}, obtaining a differential equation whose solution is precisely \eqref{potalfa1}. Note that the potential depends on the critical density $\rho_c$ and in the limit $\rho_c\rightarrow \infty$ it coincides with the potential obtained in GR. It seems clear that a potential which represents only matter degrees of freedom can in principle not contain the critical energy because it appears as a quantum geometrical effect, but if one wants to obtain an orbit which leads to the same  dynamics in LQC as a perfect fluid with EoS $P=w\rho$, i.e. $\rho(t)=\frac{\rho_c}{\frac{3}{4}(1+w)^2\rho_ct^2+1}$ \cite{Haro12}, it is mandatory to use it. We do not know the reason why $\rho_c$ appears in the potential, but what is sure is that using another potential one would obtain different orbits leading to bouncing backgrounds, none of which would mimic the one given by a perfect fluid with a linear EoS.

\subsection{Qualitative analysis in LQC}

We want to analyse the behaviour of the solution corresponding  to the  EoS $P=\omega\rho$ with $\omega>-1$. Using the potential found in \eqref{potalfa1}, we will study the dynamics of the equation
\begin{equation}
\ddot{\varphi}+3H_{\pm}(\varphi,\dot{\varphi})\dot{\varphi}+V_{\varphi}=0, \label{dinphi}
\end{equation}
where $H_{\pm}(\varphi,\dot{\varphi})=\pm \sqrt{\frac{\rho(\varphi,\dot{\varphi})}{3}\left(1-\frac{\rho(\varphi,\dot{\varphi})}{\rho_c}\right)}$,  with $\rho(\varphi,\dot{\varphi})=\frac{\dot{\varphi}^2}{2}+V(\varphi).$

\

\

Firstly, we note that from the dynamical system it follows that
\begin{equation}
\dot{\rho}=-3H_{\pm}(\varphi,\dot{\varphi})\dot{\varphi}^2.
\end{equation}

Therefore, the evolution in time will take place in an anti-clockwise sense throughout the ellipse, being $(0,0)$ a fixed point.

\

Before proceeding to analyze the different cases, we are going to take a glance at the geometry of the phase space $(\varphi,\dot{\varphi})$.

\

\begin{figure}[H]
\begin{center}
\includegraphics[height=70mm]{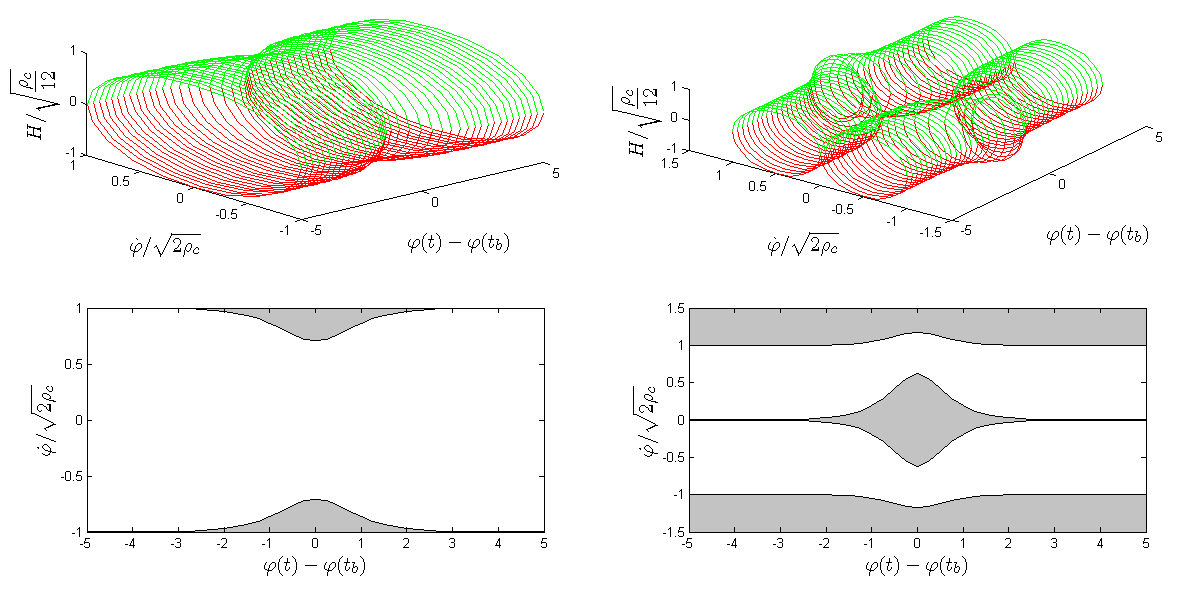}
\end{center}
\caption{Top row: the phase space $(\varphi,\dot{\varphi},H)$ for $|\omega|< 1$ (left) and $\omega>1$ (right). The green (resp. red) lines recover all the region corresponding to the expanding (resp. contracting) phase. Bottom row: its projection in the plane $(\varphi,\dot{\varphi})$, in which the white region delimited by the curves $H(\varphi,\dot{\varphi})=0$ is the allowed one.
The orbits will be plotted for $|\omega|< 1$ in Figure \ref{figwl1}, and for $\omega>1$ in Figure \ref{figwg1}.}
\label{espaifase}
\end{figure}

Since we are dealing with a bi-valued dynamical system, we need a cover 2:1 (of two sheets) of the allowed region in the plane of the phase space $(\varphi,\dot{\varphi})$, which is ramified in the curves $H(\varphi,\dot{\varphi})=0$. Hence, for the case $|\omega|<1$, this is a cylinder, whereas for $\omega>1$ we have two cylinders, as we can see in Figure \ref{espaifase}. This explains why in the phase portrait that we will later obtain we can have intersecting orbits, which happens always between an orbit in the expanding phase and another one in the contracting phase.

\

Therefore, when solving the dynamical system one option would be to use local coordinates in a cylinder. However, this would be cumbersome and, thus, we have opted for integrating the solution taking into account whether we are in the expanding or contracting phase, so that we change sign of $H$ when reaching the curve $H=0$. For the numerical results, we will use an RK78 method, that holds in memory the sign of $H$ and changes it when we switch from the contracting to the expanding phase.

\

Now, we treat separately the following cases:

\begin{itemize}

\item $\omega=1$: In this case, the potential is zero. Thus, \eqref{dinphi} becomes $\ddot{\varphi}=\mp \sqrt{\frac{3}{2}} |\dot{\varphi}|\dot{\varphi}\sqrt{1-\frac{\dot{\varphi}^2}{2\rho_c}}$.

Since $\rho\leq \rho_c$, $|\dot{\varphi}|\leq \sqrt{2\rho_c}$. Therefore, we can use the change of variables $\dot{\varphi}=\sqrt{2\rho_c}\cos(\xi)$. Assuming that at t=0 we are in the expanding phase of the semi-plane $\dot{\varphi}_0>0$, we will have $\tan\xi-\tan\xi_0=\sqrt{3\rho_c}t$. Hence,
\begin{equation}
\dot{\varphi}(t)=\sqrt{\frac{2\rho_c}{1+\left(\sqrt{3\rho_c}t+C\right)^2}} \ ,  \ \quad \mbox{for} \quad  t>\frac{-C}{\sqrt{3\rho_c}},
\end{equation}
where $C=\sqrt{\frac{2\rho_c}{\dot{\varphi}_0^2}-1}$.

It is easy to see that in the contracting phase $\left(t<\frac{-C}{\sqrt{3\rho_c}}\right)$, the expression of $\dot{\varphi}(t)$ would be exactly analogous. Moreover, if t=0 takes place during the contracting phase, the constant C would be defined as $C=-\sqrt{\frac{2\rho_c}{\dot{\varphi}_0^2}-1}$. Therefore, the orbit in the phase portrait is
\begin{equation}
(\varphi(t),\dot{\varphi}(t))=\left(\varphi_0+\sqrt{\frac{2}{3}}\ln\left(\sqrt{3\rho_c}t+C+\sqrt{1+(\sqrt{3\rho_c}t+C)^2} \right), \ \sqrt{\frac{2\rho_c}{1+\left(\sqrt{3\rho_c}t+C\right)^2}}\right),
\end{equation}
and 
in the semi-plane $\dot{\varphi}_0<0$ it is given by
\begin{equation}
(\varphi(t),\dot{\varphi}(t))=\left(\varphi_0-\sqrt{\frac{2}{3}}\ln\left(\sqrt{3\rho_c}t+C+\sqrt{1+(\sqrt{3\rho_c}t+C)^2} \right), \ -\sqrt{\frac{2\rho_c}{1+\left(\sqrt{3\rho_c}t+C\right)^2}}\right).
\end{equation}

Hence, we see that these are orbits and foliate all the space $0<|\dot{\varphi}|\leq\sqrt{2\rho_c}$, since the Equation of State is trivially fulfilled for all of them in this case, with the bounce taking place in $t_b=-\frac{C}{\sqrt{3\rho_c}}$ and such that $H(t)=(t-t_b)\rho(t)=\frac{\rho_c(t-t_b)}{1+3\rho_c(t-t_b)^2}$.

On the other hand, if $\dot{\varphi}_0=0$, the corresponding orbit $(\varphi,\dot{\varphi})=(\varphi_0,0)$ would lead to $\rho(t)=H(t)=0$.

\

Before analyzing the other cases, we  will introduce the following change of variables
\begin{equation}
    \psi =\sinh\left(\frac{\varphi\sqrt{3(1+\omega)}}{2} + \frac{1}{2}\ln\left(\frac{V_0}{2\rho_c(1-\omega)}\right)\right).
\end{equation}
Therefore,
using the potential found in \eqref{potalfa1}, equation \eqref{dinphi} becomes 
\begin{equation}
    \ddot{\psi}=-3H_{\pm}(\psi,\dot{\psi})\dot{\psi}+\rho(\psi,\dot{\psi})\psi\frac{3(1+\omega)}{2}, \label{sistema}
\end{equation}
where $H_{\pm}(\psi,\dot{\psi})=\pm\sqrt{\frac{\rho(\psi,\dot{\psi})}{3}\left(1-\frac{\rho(\psi,\dot{\psi})}{\rho_c}\right)}$ and
$\rho=\frac{2}{3(1+\omega)(1+\psi^2)}\left(\dot{\psi}^2+\frac{3(1-\omega^2)}{4}\rho_c \right)$.

\

Thus, in order to deal with this dynamical system, we compute the conditions  needed  for $\ddot{\psi}$ to vanish. They are
\begin{equation}
 \rho=0 \ \ \text{ or } \ \ \left\{  \text{sgn}(H\dot{\psi})=\text{sgn}(\psi) \text{ and } \left(|\dot{\psi}|=\frac{\sqrt{3\rho_c}}{2}(1+\omega) \text{ or } \psi^2=\frac{4}{3\rho_c(1-\omega^2)}\dot{\psi}^2 \right)\right\} .\label{condicions}
\end{equation}

Now we can proceed to analyse the rest of cases that are left:

\item $|\omega|<1$: We can distinguish two types of orbits: those that cross the axis $\psi=0$ (Type I), which physically correspond to the ones where the 
scalar field reaches the top of the potential (\ref{potalfa1}) and those that cross the axis $\dot{\psi}=0$ (Type II), which are the ones where the scalar field does not reach the top of the potential.

\

Regarding Type I orbits, we are going to consider that at the initial point $t=0$ we are at $(\psi,\dot{\psi})=(0,\dot{\psi}_0)$, where $0<\dot{\psi}_0\leq \frac{\sqrt{3\rho_c}}{2}(1+\omega)$, which comes from the restriction $0< \rho_0 \leq \rho_c$. If $\rho_0=\rho_c$, at $t=0$ we are at the bounce and, by \eqref{condicions}, the value of $\dot{\psi}$ will be the same throughout all the contracting and expanding phase, i.e., the velocity of the  scalar field does not change sign, thus reaching the top of the potential. With respect to Type II orbits, the initial point $t=0$ will be at $(\psi,\dot{\psi})=(\psi_0,0)$ where $\psi_0\neq 0$, which corresponds  to a change of sign in the velocity of the scalar field, meaning that  the orbits  do not reach the top of the potential. The corresponding phase portrait is given in Figure \ref{figwl1}, where
we have represented the set $\rho=\rho_c$, which is the discontinuous black line $\dot{\psi}=\pm \sqrt{\frac{3\rho_c}{2}(1+\omega)\left(\psi^2+\frac{1+\omega}{2}\right)}$. The pointed diagonal lines refer to the set where $\ddot{\psi}=0$, as seen in \eqref{condicions}. The blue horizontal lines are the orbits corresponding to the analytical solution.
\begin{figure}[H]
\begin{center}
\includegraphics[height=40mm]{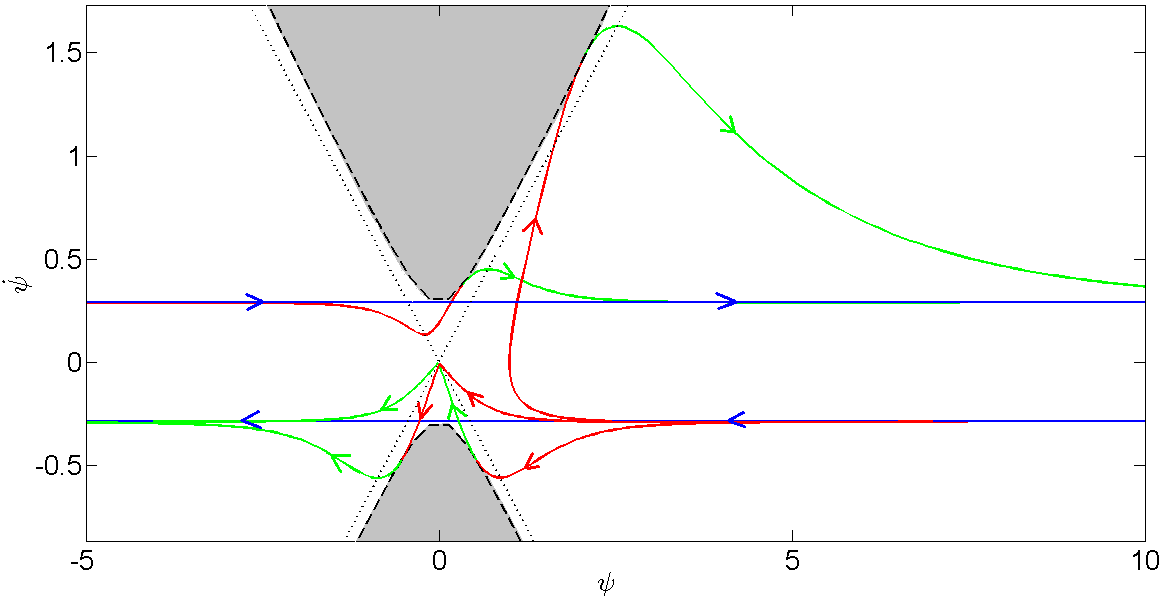}
\end{center}
\caption{Phase portrait in the phase space $(\psi,\dot\psi)$ for $w=-2/3$ and $\rho_c=1$. The blue orbits correspond to the analytic solutions which mimic the background obtained when one considers a perfect fluid with constant EoS parameter.  Since there are two dynamics, one in the contracting and another one in the expanding phase, we use two colors to draw the orbits obtained numerically: red color in the contracting and green color in the expanding phase. When the orbit reaches the bounce, i.e. when $\rho=\rho_c$, the orbit changes form red to green. Finally, the picture shows that the analytic orbits are repellers at early times and attractors at late ones.}
\label{figwl1}
\end{figure}

With respect to the rest of the curves, we have used the following colour notation: red for the contracting phase and green for the expanding phase. We have plotted one possible Type I orbit and one possible Type II orbit. In both we have considered that either $\dot{\psi}_0$ or $\psi_0$ are in the positive axis and that $t=0$ takes place during the contracting phase. We note that applying the symmetry with respect to the axis $\dot{\psi}=0$ and/or $\psi=0$ we would obtain the other possibilities for these orbits, considering that the initial point is in the negative axis and/or $t=0$ takes place during the expanding phase.

\

Finally, in Fig \ref{figwl1} we have drawn as well the invariant curves that come in and out from the saddle point (0,0), the only critical point of the dynamical system. For clarity, we have only plotted the invariant curves for $\dot{\psi}<0$. The others could be obtained with the symmetry respect to the axis $\dot{\psi}=0$. The physical meaning of these orbits is easy to understand if one takes into account the shape of the potential (\ref{potalfa1}).  Since it has a maximum,  these orbits correspond to the ones that start and end at the top of the potential. There is an orbit that ends at the top when the universe is in the contracting phase (red orbit in Figure \ref{figwl1}), one that ends at the top after the bounce (a piece red and the rest green in Figure \ref{figwl1}), another that starts at the top when the universe is in the expanding phase (green orbit in Figure \ref{figwl1}), and finally one that starts at the top when the universe is in the contracting phase and bounces to enter in the expanding one (a piece green and the rest red in Figure \ref{figwl1}).

\

So, we clearly see in each orbit the bounce at the time in which it touches the curve $\rho=\rho_c$. We observe that $\rho=0$ takes place for $\psi\to\infty$. The points in which the orbits intersect with the diagonal lines are where they change the sign of their slope. And finally the horizontal lines corresponding to the analytical solution are attractors for the expanding phase and repellers for the contracting phase.

\

We can also characterize orbits with the following magnitude:
\begin{equation}
\omega_{\textit{eff}}(t):=\frac{P(t)}{\rho(t)}=\frac{\dot{\psi}(t)^2-\frac{3}{4}(1-\omega^2)\rho_c}{\dot{\psi}(t)^2+\frac{3}{4}(1-\omega^2)\rho_c}.
\label{weff}
\end{equation}

We observe that $-1\leq\omega_{\textit{eff}}(t)<1$ and that for the analytical value $|\dot{\psi}|=\frac{\sqrt{3\rho_c}}{2}(1+\omega)$, $\omega_{\text{eff}}(t)=\omega$. Regarding the other orbits, the bounce takes place at $\omega_{\textit{eff}}(t_b)>\omega$ and, when $\rho(t)\to 0$,  since all the orbits converge to the analytic one, it can be easily verified that $\omega_{\textit{eff}}(t)\to \omega$.

\begin{figure}[H]
\begin{center}
\includegraphics[height=35mm]{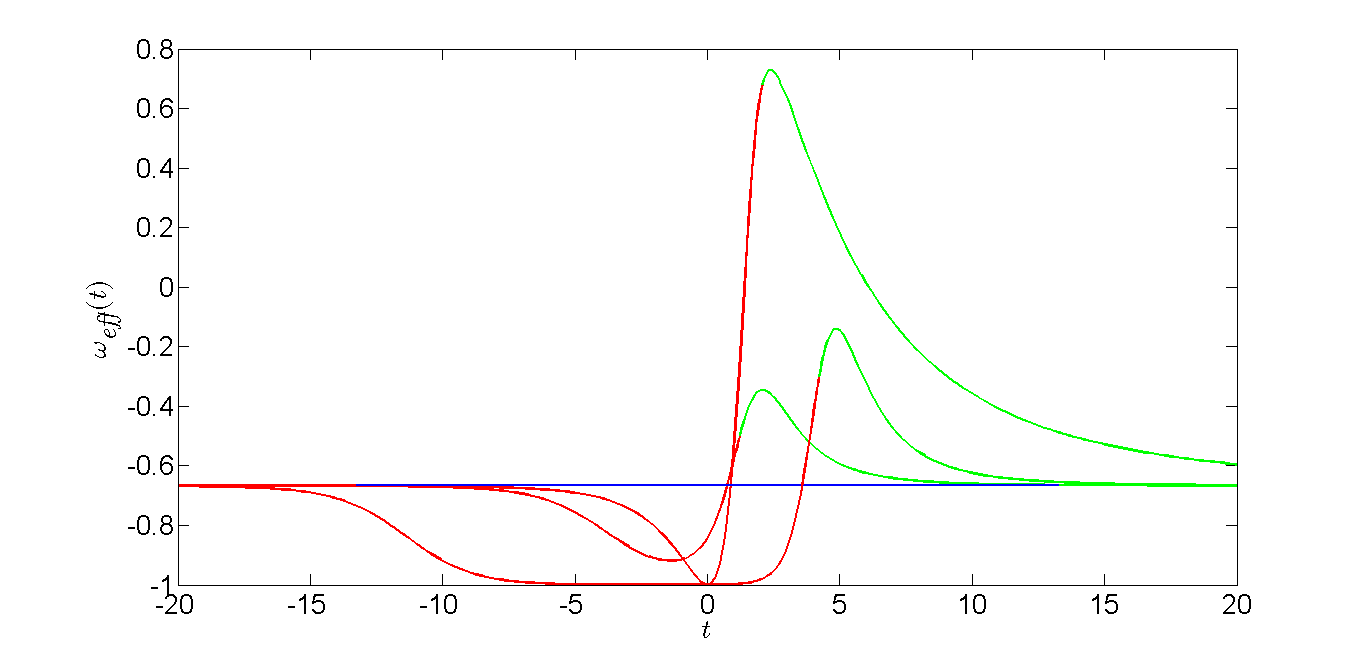}
\end{center}
\caption{Evolution of $\omega_{\textit{eff}}(t)$ for the orbits represented in the phase portrait of Figure \ref{figwl1} for $\omega=-2/3$ and $\rho_c=1$ (color code as in Fig. \ref{figwl1}).
The picture shows that at early and late times $\omega_{\textit{eff}}(t)$ converges to $\omega=-2/3$.}

\end{figure}

\item $\omega>1$: In this case, since the potential is negative, we have the lower bound of $|\dot{\psi}|\geq \frac{\sqrt{3\rho_c}}{2}\sqrt{\omega^2-1}$. Therefore, we only have Type I orbits because the velocity of the scalar field can not vanish due the Friedmann equation ($H^2$ is positive, not negative) and, thus,
there are no invariant orbits as in the previous case.

If $\rho_0=0$, we are stuck in this value of $\rho$ during all the orbit $|\dot{\psi}|=\frac{\sqrt{3\rho_c}}{2}\sqrt{\omega^2-1}$. If $\rho_0=\rho_c$, analogously as in the $|\omega|<1$ case, we stay throughout all the contracting and expanding phase in the analytical solution. 
In the phase portrait  we have represented as a discontinuous black line the curve corresponding to $\rho=\rho_c$. The other discontinuous horizontal lines that delimit the forbidden region refer to an orbit with $\rho=0$. The blue horizontal line corresponds to the orbit coming from the analytical solution.

\begin{figure}[H]
\begin{center}
\includegraphics[height=46mm]{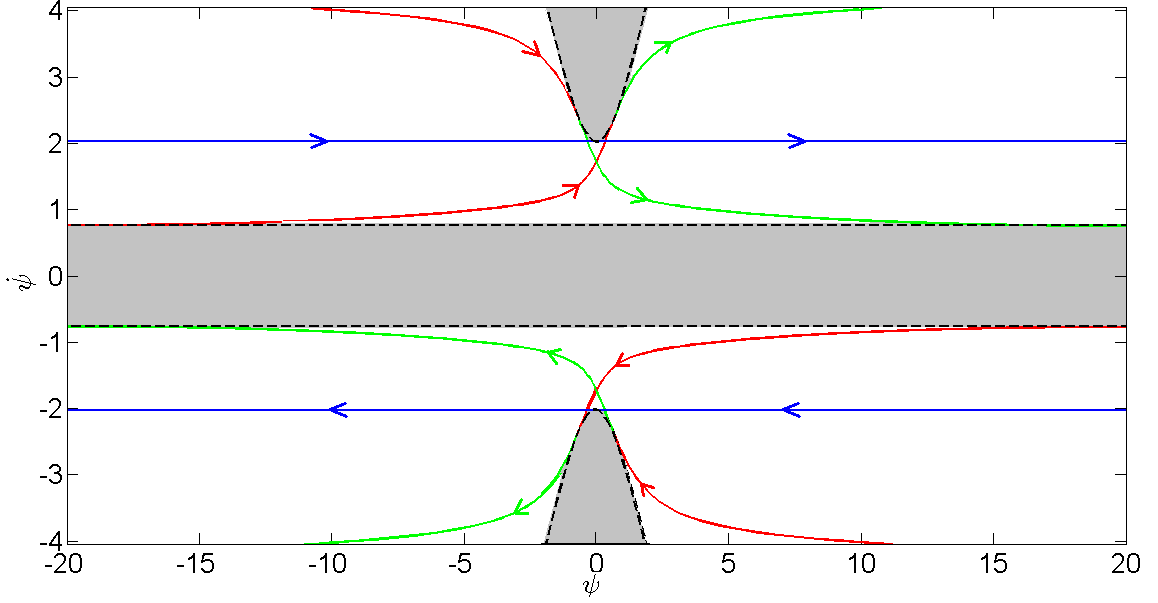}
\end{center}
\caption{Phase portrait in the phase space $(\psi,\dot\psi)$ for $w=2$ and $\rho_c=1$. We observe two bounces at $\rho=\rho_c$ and one bounce at $\rho=0$. In the first picture we observe an orbit depicting a universe in the contracting phase (red piece) which bounces and enters in the expanding one (green piece). In the second picture these orbits (green piece) enter once again to 
the contracting phase (red piece) when it reaches the black discontinuous horizontal line $\rho=0$ and finally it performs a second bounce to enter in the expanding phase (green piece).}
\label{figwg1}
\end{figure}

Regarding the other orbit, we have used the same colour notation as before.  We note that near $|\dot{\psi}|=\frac{\sqrt{3\rho_c}}{2}\sqrt{\omega^2-1}$ (corresponding to $\rho=0$), the dynamical system behaves as in General Relativity and, thus, as we already justified in this case, the orbit reaches this value in a finite time, observing a bounce for $\rho=0$. As in the case $|\omega|<1$, there is as well a bounce for $\rho=\rho_c$. 

\

 It is important to point out that, since $\omega>1$, the condition $\psi^2=\frac{4}{3\rho_c(1-\omega^2)}\dot{\psi}^2$ is never fulfilled. Hence, equation \eqref{condicions} implies that the sign of the slope of the orbit can only change when reaching $\rho=0$. This guarantees us that there will be a transition from the expanding to the contracting phase at $\rho=0$, thereby all orbits different from the analytical one depict in the $(H,\rho)$ plane a cyclic universe (it runs clockwise twice the ellipse). The beginning of the first contracting phase and the end of the second expanding phase will take place respectively for $t\to-\infty$ and $t\to\infty$, since in these cases both $\psi$ and $\dot{\psi}$ diverge in such a way that they approach $\rho=0$. Thus, we can conclude that the analytical solution is a repeller in the expanding phase and an attractor (though not global) in the contracting phase.

\

In this case, the same equation \eqref{weff} is valid. We observe that $\omega_{\textit{eff}}>1$ and that $\omega_{\textit{eff}}=\omega$ for the analytical orbit. In the other orbit, the bounces at $\rho=\rho_c$ take place at $\omega_{\textit{eff}}<\omega$ and the parameter diverges at the bounce at $\rho= 0$. Regarding $t\to-\infty$ and $t\to\infty$, the effective parameter asymptotically approaches $1$.

\begin{figure}[H]
\begin{center}
\includegraphics[height=45mm]{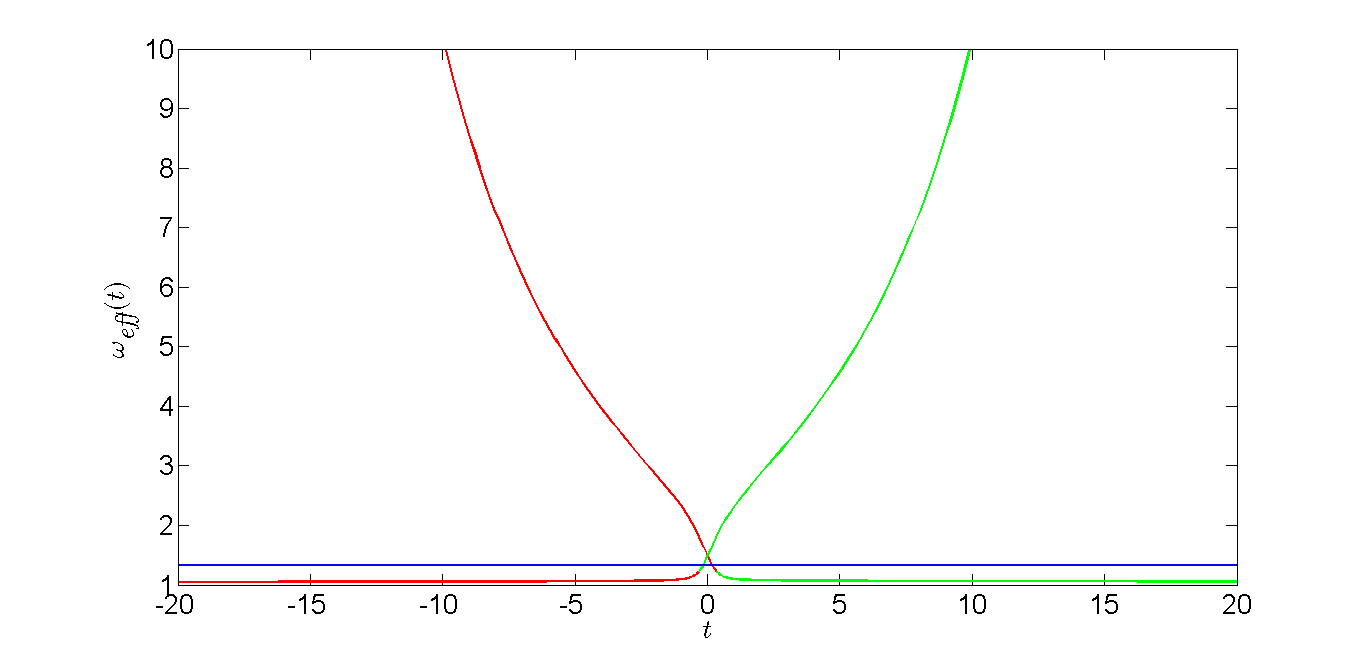}
\end{center}
\caption{Evolution of $\omega_{\textit{eff}}(t)$ for the orbits represented in the phase portrait of Figure \ref{figwg1} for $w=2$ and $\rho_c=1$ (color code as in Fig. \ref{figwg1}).
The effective EoS parameter diverges when the universe passes from the expanding to the contracting phase and it converges asymptotically to $1$ (mimicking a stiff fluid) at early and late times.
}
\label{figweffwg1}
\end{figure}

In Figure \ref{figweffwg1}, we clearly appreciate the two cycles of the orbit:

\begin{itemize}
    
    \item Cycle A: During the contracting phase the orbit comes asymptotically from $|\dot{\psi}|\to\infty$ ($\omega_{\textit{eff}}\to 1$), so that the value of $|\dot{\psi}|$ is above the one of the orbit of the analytic solution. Then it bounces, crosses this orbit and finishes the expanding phase with $|\dot{\psi}|\to \frac{\sqrt{3\rho_c}}{2}\sqrt{\omega^2-1}$, i.e. $\omega_{\textit{eff}}\to \infty$, which corresponds to the transition from the expanding to the contracting phase at $\rho=0$.

\item Cycle B: During the beginning of the contracting phase the orbit comes from $|\dot{\psi}|=\frac{\sqrt{3\rho_c}}{2}\sqrt{\omega^2-1}$ ($\omega_{\textit{eff}}\to\infty$) in such a way that the value of $|\dot{\psi}|$ is below the one of the orbit of the analytic solution. Then, it crosses this orbit, bounces and in the expanding phase it asymptotically approaches $|\dot{\psi}|\to\infty$, i.e. $\omega_{\textit{eff}}\to 1$.
\end{itemize}

\end{itemize}

\section{Conclusions}

In this manuscript, we have introduced a scalar field $\varphi$ that accounts for the perfect fluid with a linear Equation of State $P=\omega\rho$ that fills the space-time and we have used its corresponding potential $V(\varphi)$ both in GR and in LQC so as to make a qualitative study of the orbits in the phase space $(\varphi,\dot{\varphi})$, concluding that, for a canonical scalar field (i.e., $\omega>-1$), in the expanding (resp. contracting) phase, the analytical solution is an attractor (resp. repeller) for $|\omega|<1$ both in GR and LQC. For $\omega>1$, both in GR and in LQC it is a repeller (resp. attractor) in the expanding (resp. contracting) phase. However, whereas in GR the analytical solution is a global attractor in the contracting phase and a repeller in the expanding one, in LQC 
the other solutions do not catch (do not converge asymptotically) the analytical one because of the bounce, and when they enter in the expanding phase they move away from the analytical orbit, depicting at late times a universe with an effective EoS parameter equal to $1$.

\section*{Acknowledgments}

This investigation has been supported in part by MINECO (Spain), projects MTM2014-52402-C3-1-P and MTM2015-69135-P.
\

\end{document}